# AI-Driven Resource Allocation in Optical Wireless Communication Systems

Abdelrahman S. Elgamal[1], Osama Z. Aletri, Barzan A. Yosuf, Ahmad Adnan Qidan[2], Taisir El-Gorashi[2] and Jaafar M. H. Elmirghani[2]

[1]School of Electronics and Electrical Engineering, University of Leeds, Leeds, UK, LS2 9JT
[2]Department of Engineering, Kings' College London, London, United Kingdom
e-mail: el17asya@leeds.ac.uk

**ABSTRACT**

Visible light communication (VLC) is a promising solution to satisfy the extreme demands of emerging applications. VLC offers bandwidth that is orders of magnitude higher than what is offered by the radio spectrum, hence making best use of the resources is not a trivial matter. There is a growing interest to make next generation communication networks intelligent using AI based tools to automate the resource management and adapt to variations in the network automatically as opposed to conventional handcrafted schemes based on mathematical models assuming prior knowledge of the network. In this article, a reinforcement learning (RL) scheme is developed to intelligently allocate resources of an optical wireless communication (OWC) system in a HetNet environment. The main goal is to maximise the total reward of the system which is the sum rate of all users. The results of the RL scheme are compared with that of an optimization scheme that is based on Mixed Integer Linear Programming (MILP) model.

## 1. INTRODUCTION

A growing demand for high data rate communication has been witnessed in the recent years, thanks to the developments of smart terminals and infrastructures as well as a diversified set of applications such as cloud gaming, virtual and augmented reality, remote surgery and holographic projection to name a few [1]. The scale of future networks is reported to be humongous, for instance, in the near future, the number of internet-connected devices is expected to be greater than half of the earth's population which leads to an increase in global IP traffic by three folds [2]. As a result, the networks based solely on the RF spectrum will not be able to fulfil future demands for bandwidth, hence it needs to be supplemented with optical wireless communication (OWC) systems that use higher data rate frequencies offered by spectrums in the regions of infrared (IR) and visible light [3].

Visible light communication (VLC) has gained significant attention from both academia and industry for indoor applications due to the following facts: 1) The advantages that VLC offers in comparison to RF based systems; including a bandwidth of more than 1000 times the total RF spectrum [4] and robustness against interference due the large frequency bands [5] – [13], and 2) the abundance of low-cost light emitting diodes (LEDs) [14], [15].

Among other challenges, resource management arises as an important area to address to make the best use of the available resources (i.e., frequency, time, power, and wavelengths) that are offered by OWC systems. The resource management schemes in communication networks can be classified into two modes: 1) Analytical schemes based on mathematical formulations such as (i.e., convex optimization techniques) [11], and 2) artificial intelligence schemes that use machine learning algorithms to make automatic and adaptive adjustments in an intelligent way [12], [13], [16]. The main issue that arises with the use of mathematical based formulations is the fact that they require complete knowledge of the network which may not always be feasible and their complexity grows with the density of the network. The artificial intelligence schemes have garnered significant research attention due to the advent of 5G technologies, intelligent wireless terminals, and intelligent communication networks.

Taking these challenges into consideration, reinforcement learning (RL) has become a potential solution that attracted the attention of researchers in the wireless communication field. RL is a machine learning technique that relies on trial and error (similar to teaching a child) for learning the optimal behaviour through direct interaction with the environment [17]. At each time epoch, the software agent (decision maker) will take an action and observe the reward/penalty resulted from this action. In addition, the agent will observe the new state of the environment to map the behaviour of the environment in response to the action taken. This mechanism gives RL the advantage of learning optimal solutions without any prior knowledge of the environment. This feature will allow practical implementation of many optimization problems.

One of the major challenges in OWC systems is the provision of multi-user support [7], [14]. Wavelength division multiple access (WDMA) was proposed to allow access points to serve multiple users by assigning a different wavelength for each user [18] – [20]. This introduces noise and interference when multiple access points are deployed within the same room. The work in [21] introduced a Q-learning (QL) algorithm to provide resource allocation in WDMA-based VLC systems. The main objective was to allocate users to a specific wavelength in a certain access point in a way that maximised the signal to interference and noise ratio (SINR) under minimum quality of service constraints. This mechanism was compared to the non-practical mixed integer linear programming (MILP) approach and it was found that Q-learning algorithm could provide an acceptable suboptimal assignment. In a different context, the work in [22] introduced the use of QL for resource allocation in laser-based OWC systems. Multiple vertical cavity emitting lasers (VCSELs) were used to form a VCSEL array, which can jointly serve one user. The system considered in [22] comprised of four access points (APs) in which each AP consists of 4 VCSEL arrays for the purpose of multi-user support. The results demonstrate the ability of QL to practically achieve results comparable to that of the MILP model without the need for previous information or training data.

In this paper, resource allocation will be studied in an optical wireless heterogeneous network (OW HetNet) where multiple optical wireless cells co-exist in the same room using reinforcement learning (RL). The remainder of this paper is organised as follows: Section 2 gives a brief description of co-existing optical wireless HetNets. The proposed RL framework for resource allocation is discussed in Section 3. The simulation setup and results are demonstrated in Section 4. Section 5 gives the conclusions.

## 2. CO-EXISTING OPTICAL WIRELESS HETNETS

The system considered in this paper is an empty room with dimensions ($Width \times Length \times Height$). Three cells are considered to co-exist in the room namely Micro cell, Pico cell, and Atto cell. The Micro cell is located in the centre of the ceiling and has a circular coverage area of 4m. Four Angle diversity transmitters (ADTs) [23] which consist of five access points (APs) with different orientations are fixed in the four corners of the room. Each ADT provides one Pico cell with 2.6m coverage area and four Atto cells with 1.2m coverage area. Moreover, each access point can support multiple users using wavelength division multiple access (WDMA) which eliminates interference between users served by the same AP. Users are uniformly distributed on the communication plane (1m above the floor) and equipped with a single wide field of view (FOV) photodetectors to collect the optical signal as shown in Fig. 1.

The channel model and the power received by each user is calculated using the ray-tracing algorithm considering the power generated from the direct link (line of sight) and first and second order reflections [24]. Other higher order reflections will be discarded in this work as it has negligible impact on the received power and the channel characteristics [25], [26].

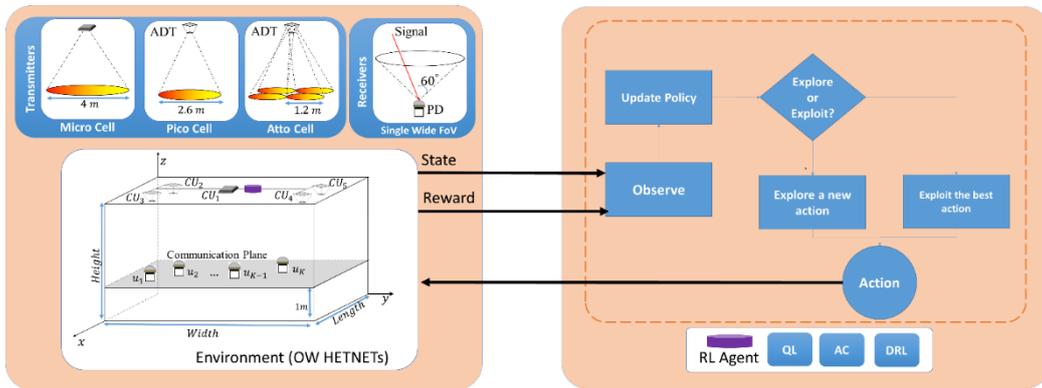

*Figure 1. RL-based System Model*

## 3. RL-DRIVEN RESOURCE ALLOCATION

To provide resource allocation using RL, several parameters need to be defined. The resource allocation problem should be formulated using Markov Decision Processes (MDP), which is a mathematical representation for stochastic problems [27] – [29]. Such problems have five main components: environment, agent, state-space, action-space, and rewards. The mapping between rewards resulting from taking each action in each specific state is called policy. The RL agent interacts with the environment to learn the optimal policy that provides the maximum reward in the long run.

In this article, the environment is the optical wireless HetNet system, and the agent is in a controller located on the ceiling responsible for the decision-making process. In the state-space, each state is described by a binary vector that has a length equal to the number of users. Each variable in the vector determines the achievement of minimum quality of service for a specific user. If the minimum quality of service is met for a certain user, its associated binary variable in the vector will be set to 1, otherwise it will be set to 0. The action space comprises of a set of all possible permutations of allocations. Each allocation is described using a 3-D binary matrix which maps users to their associated APs and wavelengths called the assignment matrix. Each user, AP and WL has an index in the assignment matrix in which if user $u$ is to be assigned to wavelength $w$ in AP $l$, the associated variable with index $u, l, w$ in the assignment matrix will equal to 1, otherwise it will equal 0.

We consider two performance metrics: SINR and achievable data rate. Therefore, two formulations for the rewards have been introduced based on these performance matrices. In both formulations, the agent is rewarded when a higher SINR/achievable rate is achieved and penalized otherwise. The main purpose of the penalty is to emphasize and distinguish the best resource allocation in the environment.

The exploration/exploitation trade-off is a critical factor in the decision-making process; hence we consider the $\epsilon$ $-$greedy algorithm for making the exploration/exploitation decision [21], [22]. In QL, the quality of actions is measured according to the action value function (Q-function). The output of this function is stored in a Q-table that maps the quality of each action taken in each possible state from the network, which is referred to as the Q-value. This process will be repeated until an optimal policy is achieved. This policy will have the required knowledge to take actions and the adaptability to network changes.

## 4. SIMULATION SETUP AND RESULTS

In the simulation we consider a room with $4m \times 4m \times 3m$ dimensions. For simplicity, we considered only Pico and Atto cells provided by an ADT to serve eight users which are uniformly distributed below the transmitter. The ADT consists of 5 access points each covers one cell. Out of the five cells, one is a Pico cell and four are Atto cells. Each cell transmitter has different orientation (Azimuth and Elevation angles) to allow coverage of different areas in the room. A single cell can support a maximum of two users over red and yellow wavelengths. The parameters of the room, cells, access points, and users are listed in Table 1.

*Table 1. Simulation Parameters.*

| Parameter | Configuration | | | | |
|---|---|---|---|---|---|
| Walls and ceiling reflection coefficient, Floor reflection coefficient | 0.8, 0.3 [13] | | | | |
| Reflections order | 1 | | 2 | | |
| Area of reflection element (cm × cm) | 5 × 5 | | 20 × 20 | | |
| Order of Lambertian pattern, walls, floor and ceiling | 1 [13] | | | | |
| Semi-angle at half power of Pico cell | 42º | | | | |
| Semi-angle at half power of Atto Cell | 22º | | | | |
| Number of RY LDs per Cell | 6 | | | | |
| Transmitted optical power of Red (R), Yellow (Y) wavelengths respectively. | 0.8, 0.5 W | | | | |
| Transmitters location (x, y, z) | (1m, 1m, 3m) | | | | |
| **Transmitter ID** | **Pico** | **Atto#1** | **Atto#2** | **Atto #3** | **Atto #4** |
| Transmitter Orientation (Azimuth) | 0º | 45º | 135º | 225º | 315º |
| Transmitter Orientation (Elevartion) | -90º | -65º | -65º | -65º | -65º |

The assignment of users, cells, and access points was done using QL and was compared to the optimal solution of MILP. First, the allocation process was done with an objective function of maximising the SINR. During the learning process, the RL agent managed to find an acceptable allocation under SINR maximisation. This allocation provides a suboptimal total SINR very close to the MILP as shown in Fig.2a. Leaving the agent more in the learning phase, has led the RL agent to learn the optimal allocation. This optimisation problem was found to have multiple optimal allocations, hence, the optimal allocation provided by QL is slightly different from the one provided by MILP in terms of users 6 and 8. The SINR for these users are different in both solutions whilst the overall behaviour is similar and a minimum SINR of 36 is maintained to guarantee QoS. The total SINR for the solution provided by QL is similar to the optimal solution provided using MILP as demonstrated in Fig. 2c.

The RL allocation process has been developed to maximise the achievable rate in $bits/s/Hz$. Several solutions have been provided by QL that can achieve the same total rate achieved by the MILP as shown in Fig. 2b. The minimum rate of 8 $bits/s/Hz$ has been achieved for all users. Therefore, the QL agent was able to find multiple optimal solutions when rate maximisation is the objective function as shown in Fig.2d.

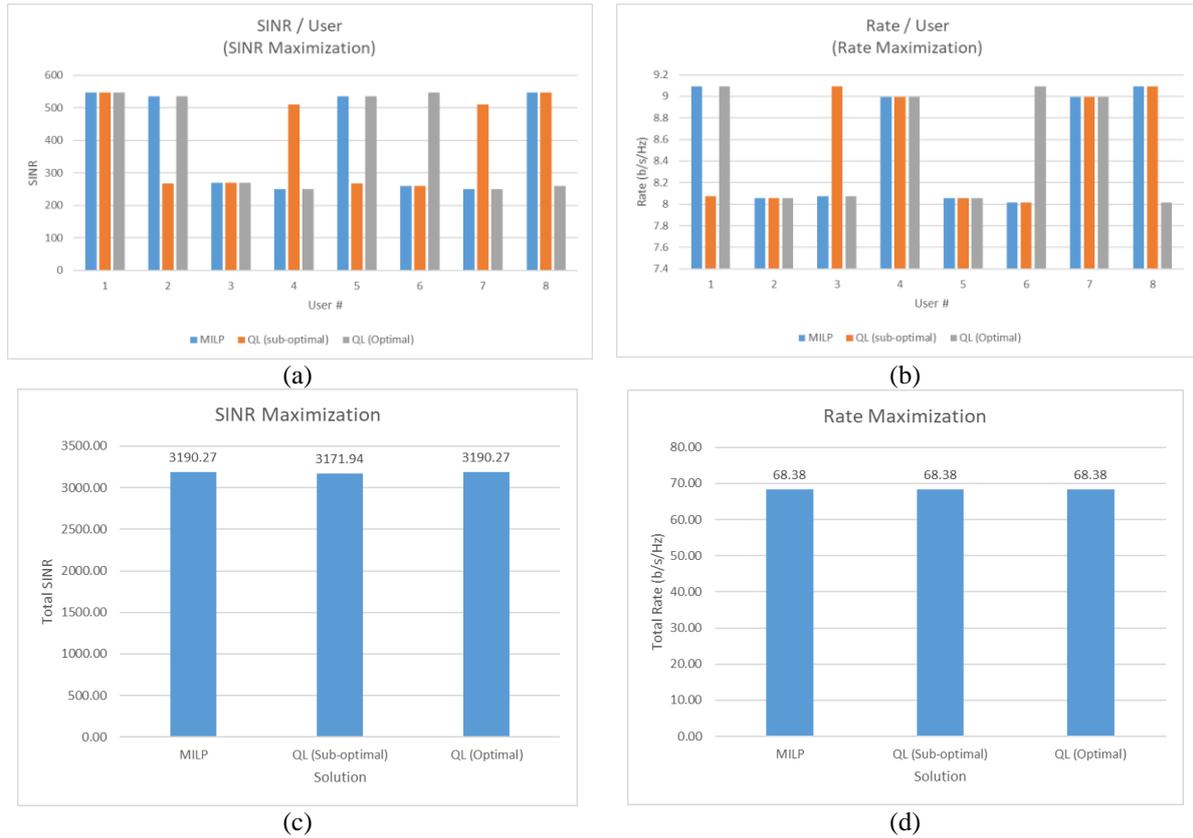

*Figure 2. (a) SINR per User (b) Achievable Rate per User (c) Maximised Total SINR (d) Maximised Achievable Rate*

## 5. CONCLUSION

**6.** This paper has developed a RL based resource allocation scheme to intelligently allocate the resources of an optical wireless system in a HetNet environment. We examine RL with two objectives: 1) total SINR maximisation. 2) sum rate maximisation. The same resource allocation problem was formulated using mixed integer linear programming (MILP) for comparison purposes. The results showed that the RL approach managed to achieve comparable results to the optimal results of the MILP.

## ACKNOWLEDGEMENTS

This work has been supported in part by the Engineering and Physical Sciences Research Council (EPSRC), in part by the INTERNET project under Grant EP/H040536/1, and in part by the STAR project under Grant EP/K016873/1 and in part by the TOWS project under Grant EP/S016570/1. All data are provided in full in the results section of this paper. ASE author would like to acknowledge EPSRC for funding his PhD scholarship.